\documentclass[letterpaper,conference]{IEEEtran}
\usepackage[utf8]{inputenc}
\usepackage{latexsym}
\usepackage{ngerman}
\renewcommand{\figurename}{Figure}
\renewcommand{\tablename}{Table}
\renewcommand{\abstractname}{Abstract}
\usepackage{graphicx}
\usepackage{array}
\usepackage{multirow}
\usepackage{hhline}		
\usepackage{subfigure}
\usepackage{mathptmx}
\usepackage{mathtools}
\usepackage{amsmath}
\usepackage{amsfonts}
\usepackage{amssymb}
\usepackage{amsbsy}
\usepackage{amsthm}
\usepackage{todonotes}
\usepackage{listings}
\usepackage{color} 
\usepackage{float}
\usepackage[justification=centering]{caption}
\usepackage[export]{adjustbox}
\usepackage{cite}

\begin{document}
\lstset{                          
language = Python,                
numbers = left,                   
numberstyle = \footnotesize,      
stepnumber = 5,                   
numbersep = 5pt,                  
backgroundcolor = \color{white},  
showspaces = false,               
showstringspaces = false,         
showtabs = false,                 
frame = single,                     
tabsize = 2,                       
captionpos = b,                   
breaklines = true,                
breakatwhitespace = false        
} 

\title{Tracemax: A Novel Single Packet IP Traceback Strategy for Data-Flow Analysis\vspace*{-2mm}}
\author{
\IEEEauthorblockN{Peter Hillmann, Frank Tietze, and Gabi Dreo Rodosek}
\IEEEauthorblockA{Universit\"at der Bundeswehr M\"unchen\\
Neubiberg, 85577, GERMANY\\
Email: \{peter.hillmann, frank.tietze, gabi.dreo\}@unibw.de
}
}

\maketitle


\begin{abstract}
The identification of the exact path that packets are routed on in the network is quite a challenge. This paper presents a novel, efficient traceback strategy named \textit{Tracemax} in context of a defense system against distributed denial of service (DDoS) attacks. A single packet can be directly traced over many more hops than the  current existing techniques allow. In combination with a defense system it differentiates between multiple connections. It aims to letting non-malicious connections pass while bad ones get thwarted. The novel concept allows detailed analyses of the traffic and the transmission path through the network. The strategy can effectively reduce the effect of common bandwidth and resource consumption attacks, foster early warning and prevention as well as higher the availability of the network services for the wanted customers.
\end{abstract}
\IEEEpeerreviewmaketitle

\renewcommand\IEEEkeywordsname{Keywords}
\begin{IEEEkeywords}
Computer network management, IP networks, IP packet, Traceback, Packet trace, Denial of Service
\end{IEEEkeywords}
\section{Introduction}\label{introduction}
One of the most publicly recognized distributed denial of service (DDoS) attacks took place in 2010. As protest reaction on inhibition of WikiLeaks bank accounts, MasterCard, Visa and PayPal as well as other large finance institutes got attacked \cite{forbes10}.
The coordinated action named \glqq Operation Payback\grqq{} has stopped aforesaid services for several hours causing great financial loss.
In 2014, the largest detected DDoS attack had consumed a bandwidth of over 500 Gbit/s\cite{forbes14}. Nowadays, no service infrastructure can handle such large amount of data. 
%
The power and danger of DDoS attacks is an increasingly frequent problem in the global Internet. Attacks are not only focused against single services but entire countries and infrastructures.
An adequate traceback and defense strategy is needed in order to identify multiple sources and to reveal spoofed IP addresses. 
It is important to get provable identification and localization of the attackers for law enforcement and forensic analysis. Of large interest for states is especially the source of military cyber operations or protest groups. 


\section{Scenario}\label{scenario}
\begin{figure}[htbp]
 {
 \vspace{-3mm}
 \centering
 \includegraphics[width=0.34\textwidth]{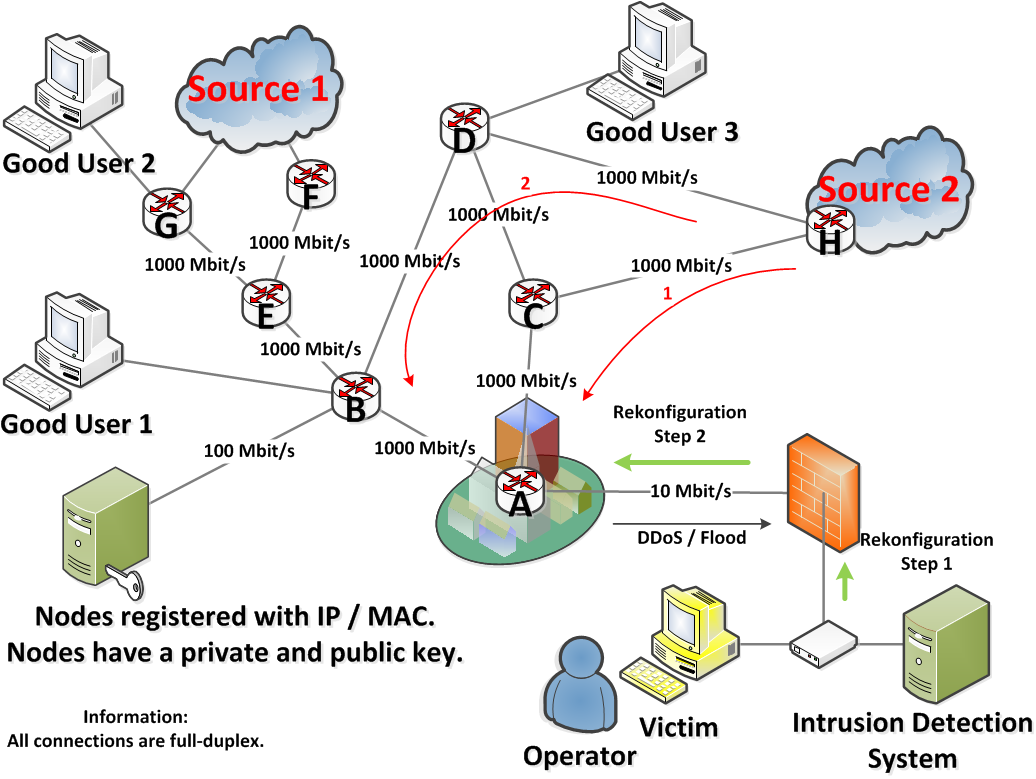}
 \caption{Traceback of sophisticated attackers.}
\label{TracebackAttacker}
\vspace*{-4mm}
 }
 \end{figure}

The idea of an traceback strategy provided by an Internet service provider (ISP) is illustrated by using the following real world scenario, see Figure \ref{TracebackAttacker}. A DDoS attack and malicious traffic are detected at a terminal system by an intrusion detection systems (IDS). 
The associated alarm vector triggers the labeling service at the victims own edge router. It propagates self-organizing this labeling task through the network. The service is secured by cryptographic authentication. After the configuration of the routers, all messages can be traced back. The traffic data, path labeling data and attack meta data are prepared and stored with time stamp to cope with the forensic needs for trustworthy and transparent path reconstruction of law enforcement. The service extends the data for digital forensics in a live response scenario. 
In parallel, the system starts defense actions by propagating the attack signature to the routers. It is used for filtering, blocking or delaying of corresponding incoming traffic and reduces the network load.
%
%

Other use cases are the verification of traffic shaping configured in the network management system and identification of hidden channels. Load balancing and zone routing  can be verified by packet tracking as well. In addition, the technique provides an approach to detect Cormelt attacks with currently no defense mechanisms existing. Thereby, a Botnet generate pseudo realistic traffic between the bots at a specific network node and overload it. 


\section{Requirements and assumptions}\label{sec:requirements}
For our analysis, we assume that routes can dynamically change during an attack. Furthermore, packets can get lost, and the order of the packets within a connection can be changed. Finally, attackers are able to generate any packet with any faked parameters. Attackers are able to cooperate and have knowledge of traceback strategies. According to the scenario and the various application areas, we address traceback strategies concerning to the following main aspects:


\begin{itemize}
\vspace*{1mm}
\item \textbf{Single packet traceback} 
\item Detect and Differentiate \textbf{multiple attackers}
\item \textbf{Fast path reconstruction} and \textbf{preventive applicable}
\item \textbf{Traced Hops / Locations} of more than 50 hops\\ 
\vspace*{-3mm}
\end{itemize}
\newpage
A call of the website \textit{''torproject.org''} from our location is transferred over more than 18 routers. It does not include proxy servers or anonymizing techniques. The worst case number of nodes traversed by a packet was about 56 hops in 2011 \cite{Chinnery2010}.


\section{Related work}\label{relatedwork}
Over the past years, many traceback and defense strategies have been developed.
The approach of Deterministic Router Stamping (DRS)\cite{25} is only able to trace up to 9 hops whereas Probabilistic Router Stamping (PRS) need many packets to reconstruct the path.
Packet Marking (PM) \cite{27} is differentiated in Node-Sampling and Edge-Sampling. They are not efficient with respect to the capacity. The marking information in the payload can cause errors. 
Link Testing by Input Debugging (LTID) \cite{29} and Link Testing by Controlled Flooding (LTCF) \cite{30} require a continuous attack for complete attack path identification and limit immediate live defense actions. 
%
%
%
The approach of Logging at a router \cite{35} creates a large amount of data, needs a lot resources and limit live response actions.
%
%
The backward discovered path of the Internet Control Message Protocol (ICMP) traceback approach \cite{36} does not necessarily match the real path of a packet, because of load balancing and other influences.
The ISP Traceback \cite{stelte13, Khan} 
 allows the identification of the source ISP, but not track the direct path. There are more algorithms and hybrid solutions to trace packets as in \cite{Gong2008}. 
%

All in all, the known traceback strategies do not fulfill the identified requirements to trace the path of network packets. 

\section{The new Approach: Tracemax}
In reference to the scenario in Section \ref{scenario}, a user sends a specific request to his ISP to start labeling packets. 
This activates the novel traceback algorithm \textit{Tracemax} for a predefined number of routers. These routers have to cooperate and have to be configured for it. 
\textit{Tracemax} consists mainly of a marking scheme and a reconstruction method. The routers are marking packets on the path during the transmission. The reconstruction method determines the path of a packet afterwards. 

\subsection{Pre-configuration and ID assignment}
Before the \textit{Tracemax} system is rolled out, every physical port $n$ of every router $T_{i}$ gets his own \textit{Tracemax} $ID_{i,n}$. An $ID$ is not necessarily unique in the entire system. Still the $ID$ numbers have to defined in a way that the path reconstruction is uniquely possible. In analogy of the logical port number, an assigned \textit{Tracemax} $ID$ is oriented on it. But for every router is not allowed to get the same incoming ID of several connected routers. The generation of the IDs is still easy under this condition. An automatic algorithm is described in \mbox{Listing \ref{lst:IDGeneration}}. 
The algorithm prevents that a preselected router gets the same ID on direct links from two neighboring routers. This is achieved by incrementing one of the intended IDs. 

\lstset{ %
  backgroundcolor=\color{white},   
  basicstyle=\footnotesize,        
  frame=single,                    
 keywordstyle=\footnotesize,
 keywords={},
 deletekeywords={all},
  numbers=none,                    
  numbersep=5pt,                   
  stepnumber=1,                    
 tabsize=2,                       
}
\lstset{aboveskip=10pt}
\begin{lstlisting}[frame=single,caption={Algorithm for automatic ID assignment.},label=lst:IDGeneration,float=htbp, belowskip=-2.2 \baselineskip]
1. Select randomly the first node.
2. Assign every port the ID as their port number n.
3. do {
4. 		selectedNode = a connected node to the already
					 selected node(s), which has not selected.
5. 		for ( all free ports of selectedNode ) {
6. 			  isAssigned = false;
7. 			  ID = 1;
8. 			  do {
9. 			    if ( ID is unused at selectedNode ) {
10. 				   if ( ID != all incoming IDs at the 
						   connected node){
11. 				      selectedNode.portID = ID;
12.		 			      isAssigned = true;
13. 				   }
14. 			  }
15. 		    ID++;
16. 		  } until ( !isAssigned )
17. 	 }
18. } until ( all nodes are selected )
\end{lstlisting}

\begin{figure}[hbtp]
\vspace*{-2mm}
     		\begin{minipage} [hbt]{4.22cm}
\begin{center}
\includegraphics[width=0.9\textwidth]{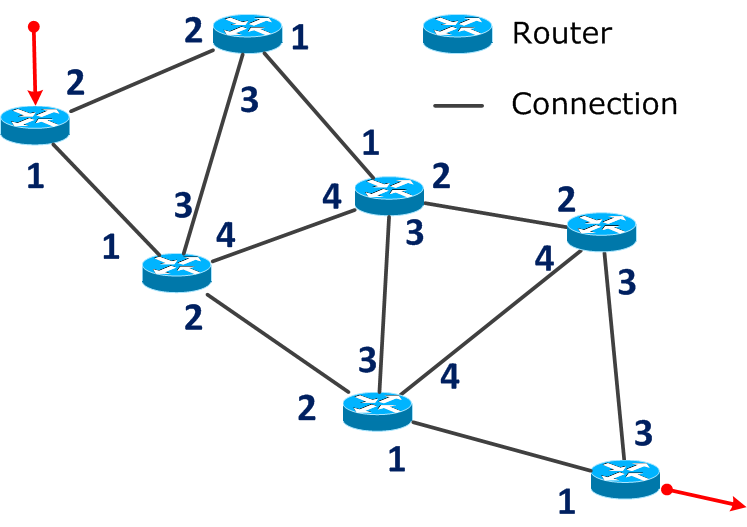}\\
\caption{Valid configuration.}
\label{Tracemax}
\end{center}
 				\end{minipage}
 					\textbf{or}\hfill
 				\begin{minipage} [hbt]{4.22cm}
\begin{center}
\includegraphics[width=0.9\textwidth]{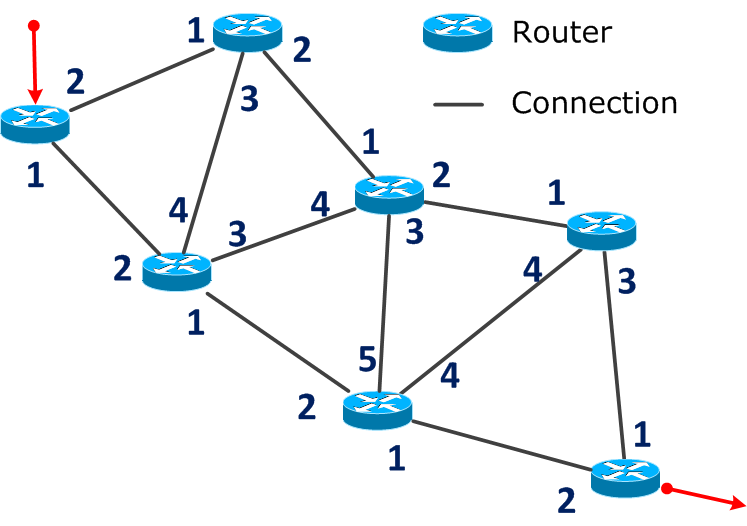}\\
\caption{Valid configuration.}
\label{Tracemax2}
\end{center}
 			\end{minipage}
 			\vspace*{-3mm}
 			\end{figure}

\begin{figure}[hbtp]
 			\vspace*{-2mm}
     		\begin{minipage} [hbt]{4.7cm}
\begin{center}
  \includegraphics[width=0.78\textwidth]{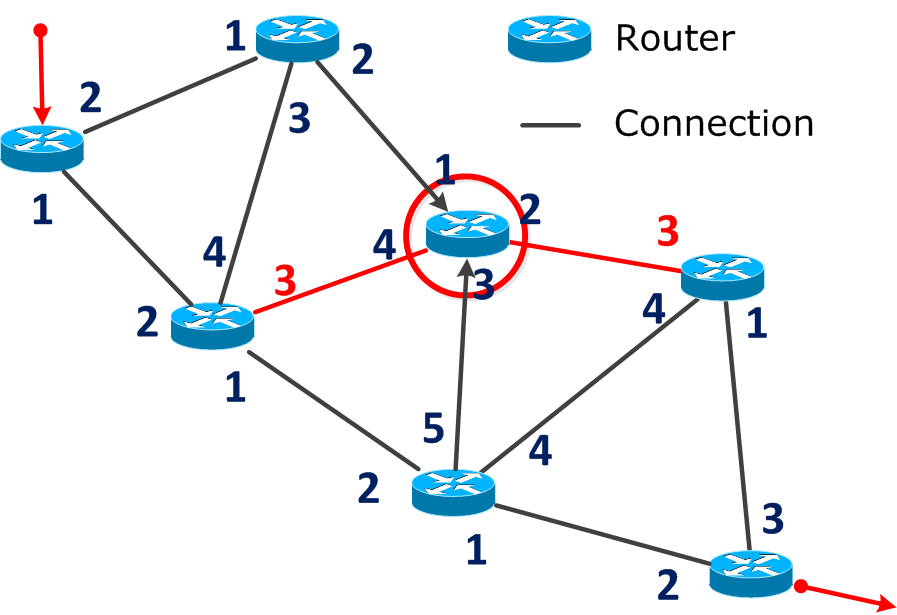}
  \caption{Invalid ID assignment.}
  \label{Tracemaxfail}
\end{center}
 				\end{minipage}
\hfill
 				\begin{minipage} [hbt]{3.9cm}
\vspace*{3mm}
\begin{center}
 \includegraphics[width=0.75\textwidth]{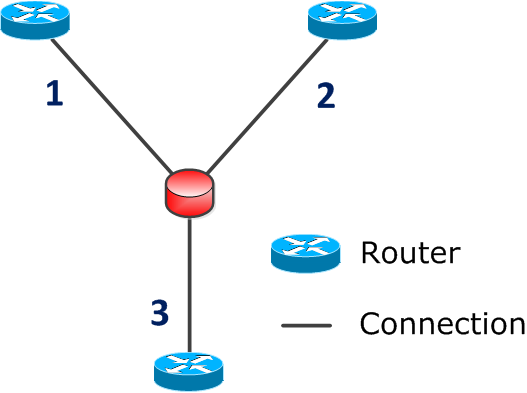}
 \caption{A bridged router.}
 \label{szenarioBridge}
\end{center}
 			\end{minipage}
 			\vspace*{-5mm}
 			\end{figure}

Two possible valid ID assignments with minimal bit size and with an automatic assigned version can be seen in \mbox{Figure \ref{Tracemax}} and \ref{Tracemax2}. An invalid example of ID assignment is shown in \mbox{Figure \ref{Tracemaxfail}}. The encircled router has twice the incoming ID number 3. Thereby the simple reconstruction function is not able to decide between the two routers as source of a packet.

A strong reconstruction algorithm, which gives attention on the transfer direction of a packet, is possible to resolve even some conflicted IDs. 
Nevertheless, every router in the network should mark packets. This avoids unclear or incomplete ID sequences, which can not be reconstructed to a transmission path. But it is not mandatory for a clear path reconstruction that the entire network is completely configured to mark packets. With genial predefined and assigned IDs single routers can be bridged, see \mbox{Figure \ref{szenarioBridge}}. This ID assignment allows a unique path reconstruction even the network device in the middle do not marking transfered packets.

\subsection{Marking scheme}
Every router $T_{i}$ on the path writes the predefined $ID_{i,n}$ in the \textit{Option Field} of the IP packet. Even if the size is very limited (40 bytes), the \textit{Option Field} in our case is best choice and the size is sufficient. This increases the network load only slightly and has no side effects. Our defined unique $ID_{i,n}$ number with less than 6 bits for labeling every port $n$ of an active network device $i$ is much smaller than 32 bits of an IPv4 address. In this way we can store many more hops in form of $ID$s into the \textit{Option Field}. The precise necessary bit size of the $ID$ number ($n = 1..k$) depends on the situation. The bit size hangs indirectly on the router $T_{i}$ with most physical connections $k$, which is in the predefined area of \textit{Tracemax}. For performance improvement, single network devices with many ports can be virtually split in multiple devices to lower the maximum necessary bit size for an $ID$ number. The $ID_{i,n}$ can be seen as a lower ISO/OSI layer information, for example the port number of a physical connection on a router. In the process of development, the \textit{Tracemax} $ID$ is declared as an abstract number, which is independent from the port number. This allows improvements during the $ID$ assignment and leads to a reduced bit size for the $ID$s. Every router $T_{i}$ writes the assigned and predefined $ID$ of the physical port into the packets \textit{Option Field} at the outgoing interface. The next router can prove the value.

\subsection{Reconstruction scheme}
The reconstruction function extracts the IDs out of the IP header. The sequence of IDs defines the used connections between the routers and thereby the path of the packet. The function needs information about the endpoint where the packet is captured. This information can be stored with the packets or the user already knows it. In most cases the reconstruction is running at the same system as the endpoint. Otherwise, the last and closest IP address to the destination could be stored completely into the \textit{Option Field} as well.
The function correlates the sequence of the IDs with the knowledge of the network infrastructure and the predefined IDs. This is done step-by-step backwards, starting from the receiver node until the last marked node is found. Due to the precise ID assignment only one possible connection is valid in every step of the backward path reconstruction. The administrator of \textit{Tracemax} has all necessary information allowing the complete reconstruction of the path. The function can also map these routers to their IP addresses. In the end, the order of the routers and the IP addresses of the path are known.

\subsection{Additional extensions}

If an incoming packet is entered into the managed \textit{Tracemax} system, the information in the \textit{Option Field} has to be deleted. This prevents injecting of false information and generates space for the own traceback information. This is the case at the borders of different \textit{Tracemax} systems or ISPs. 
An incoming packet at a \textit{Tracemax} system edge router should be labeled with the sending router IP address from the routing table or with an additional ID. Thus the path reconstruction is extended with reference to the first external routing device. 
Also if a packet leaves the managed \textit{Tracemax} components the last marking edge router can store an information about the next network device without marking functionality into the \textit{Option Field}. This also supports to bridge network devices.

In summary, the \textit{Option Field} includes in the complete version the following information:
%
[Preamble : IP-Sender : $1^{st}$ ID : $2^{nd}$ ID : $3^{rd}$ ID : ..... : n-1. ID : n. ID : IP-Receiver]. 

A stepwise roll out or limited deployment of the \textit{Tracemax} System is possible. 
Thus it allows to trace the path of a transfered packet only partly. 

A large advantage of \textit{Tracemax} is that it does not provide any information about the private topology information of an ISP even if someone gets access to the IDs in the \textit{Option Field}. The IDs could also be changed in a short interval to avoid reverse engineering. In the scenario of DDoS we have to assume that IP spoofing is used by the attacker. This aspect does not influence our \textit{Tracemax} strategy because it is not based on IP source addresses. 

\subsection{Storing the information in the IP header}\label{Option-Field}
In the following, the Preamble and the entire \textit{Option Field} is explained in more detail, because not all of the available bytes are free and flexible usable. The IP header in version 4 is 20 bytes large and offers an additional \textit{Option Field} with the variable length of up to 40 bytes, dependent on Padding.



There are two different formatting styles to add information in the \textit{Option Field}
. The traceback information can either be added as single octet. \textit{Tracemax} on the other hand uses the defined option-length to be able to predefine the necessary space.
The Preamble of all option-type parts is: 1 bit \textit{Copied Flag}, 2 bits \textit{Option Class}, and 5 bits \textit{Option Number}.

The 1 bit \textit{Copied Flag} defines that this option part is copied into all fragments. Thus every single packet is traced so it will be set to $0_{2}$.
For the 2 bit \textit{Class Field} exists different valid possibilities. 
The measurement parameter with $10_{2}$ is most fitting for our case.
A list of all different specified 5 bits \textit{Option Number} parameters can be found at \cite{ipparameters}. We have to find a value, which do not create conflicts.  \textit{Tracemax} uses the unassigned option number $1 0110_{2}$ ($22_{10}$). In combination, the $1^{st}$ octet is: $86_{10}$ = 0x56 = $0 1 0 1 0 1 1 0_{2}$

The \textit{Option Number} parameters has a \textit{traceroute option}, but it offers only the possibility to count the hops and do not allow to traceback a packet.
%
It is very important to give attention on this formatting, because of possible malformed packets or the unintentionally mapping with some other option values. For example, the two option values \textit{Loose Source Route} and \textit{Strict Source Route} are discouraged because of security concerns\cite{faqs14} and will be dropped. 

 
Furthermore, this octet is followed by the \textit{option-length} octet, which specifies the length of the current option part including the option part header
. The size can be between 0 and 40 byte. 
 It is possible to extend the header according to the demand.
 %
%
The $2^{nd}$ octet is: $40_{10}$ = 0x28 = $0 0 1 0 1 0 0 0_{2}$


This concept can be easily adapted to IP version 6. For this, it is necessary to define a new \textit{Next Header} to store the traceback information.

\vspace{1mm}
\section{Proof of concept and evaluation}\label{simulation}
The evaluation of the \textit{Tracemax} is based on experiments using a prototypic implementation. Our approach is implemented with the interactive packet manipulation and generation tool \textit{Scapy}, which is written in Python. We have set up trial installations with multiple network components and several computer. The network components for packet marking are realized by computer with appropriate router configuration. Thus, the tests are very realistic. 
%
The Python Script, Listing \ref{lst:python}, describes the marking algorithm of \textit{Tracemax}, which is used at every router. 
First, the script detects whether the incoming packet has an \textit{Option Field} or not. Accordingly, the \textit{Option Field} is copied or added with the Preamble '\textbackslash x56\textbackslash x28'. It extends the \textit{Option Field} with its specific ID defined in the variable 'myindex', here the value '\textbackslash x12'. For simplification, every router has only one ID to mark the packets with. 
We also avoid a dynamic extension of the \textit{Option Field} for the tests. After a new packet is generated, it will be send to the next network hop. For the experiments, the Script filters on ICMP packets.\\

\lstset{aboveskip=-8pt}
\lstset{language=Python} 
\lstset{ %
  backgroundcolor=\color{white},   
  basicstyle=\footnotesize,        
  frame=single,                    
  numbers=none,                    
  numbersep=5pt,                   
  stepnumber=1,                    
}
\begin{lstlisting}[frame=single,caption={Python script for Tracemax.},label=lst:python,float=htbp, belowskip=-1.4 \baselineskip]
from scapy.all import *
def chgSend(x):
	myindex = '\x12'	# Assigned Port Identifier
	
	optionsarray = x[IP].options 	#Option Field
	if optionsarray.count(1) ==  0   and
	   str(optionsarray)     != '[]' :
				optionsstring = str(optionsarray[0])
	else: optionsstring = '\x56\x28'

	#Generate packet
	y=IP(src=x[IP].src,dst=x[IP].dst,len=60,options=
		IPOption(optionsstring+myindex))/x[IP].payload
	send(y)

while 1: # Listen to the network interface
	sniff(prn=chgSend, lfilter=lambda x:
	x.haslayer(ICMP), count=1)
\end{lstlisting}


During the evaluation, every machine runs \textit{Wireshark} for packet capturing. 
The reconstruction of the path is done manually to examine and resolve possible problems. The \textit{Wireshark} traces show the packet marking in the \textit{Option Field} after each hop. 
The transmission path can be reconstructed in combination with the known information about the ID assignment. The experiments validate our concept and show the easy implementation as well as the practicability.
%
%
%
%
%
To extend the IP-Header with the entire 40 bytes, it creates a small additional network load of \mbox{2.$\overline{\text{6}}$ \%} for a 1500 bytes packet. 

To summarize the results, Table \ref{tab:comparison} compares our \textit{Tracemax} system with the existing strategies. 
The symbols have the following meaning in comparison to the other traceback techniques: + advantage; - disadvantage; o neutral.


\vspace{-1mm}
\begin{table}[htb]
\centering
\caption{Comparison of different traceback strategies}
\setlength{\tabcolsep}{1pt}
\small
\begin{tabular}{|c!{\vrule width 0.8pt}c|c|c|c|c|c|c|c|c|c|c|c|c|}
\hline
\rotatebox{90}{\textbf{Algorithms}} & \rotatebox{90}{\parbox{2.1cm}{\textbf{Traced Hops /\ \\ Locations}}} & \rotatebox{90}{\textbf{Effectiveness}} & \rotatebox{90}{\textbf{Efficiency}} & \rotatebox{90}{\textbf{Robustness}} & \rotatebox{90}{\textbf{Scalability}} & \rotatebox{90}{\textbf{Costs}} & \rotatebox{90}{\parbox{1.9cm}{\textbf{No ISP\\ cooperation}}} & \rotatebox{90}{\parbox{1.9cm}{\textbf{Deploy entire\\ Sub-Network}}} & \rotatebox{90}{\parbox{2.2cm}{\textbf{Short Detection\\ Time}}} & \rotatebox{90}{\parbox{1.9cm}{\textbf{Preventive\\ usable}}} & \rotatebox{90}{\parbox{1.9cm}{\textbf{Single Packet\\ Traceback}}} & \rotatebox{90}{\parbox{2.1cm}{\textbf{Detect multiple\\ senders}}} \\
\hline
\hhline{-------------}
Ingress Filtering & 0 & - & + & + & +  & + & - & + & + & + & + & + \\\hline
RS-DRS & 9 & +  & o & o & o & + & + & o & + & + & + & + \\\hline
RS-PRS & $\infty$ & + & o & o & o & + & + & o & - & + & - & - \\\hline
PM-Node-Samp. & $\infty$ & + & + & o & + & o & +  & o & o & + & + & + \\\hline
PM-Edge-Samp. & 1 edge & o & - & o & + & o & +  & o & - & + & - & - \\\hline
LT-Input Debug. & $\infty$ & o & - & + & o & o & o & - & - & - & - & - \\ \hline
LT-Cont. Flood. & $\infty$ & - & - & o & - & + & o  & + & - & - & - & - \\ \hline
Logging & $\infty$ & + & + & + & + & - & -  & + & - & + & + & + \\ \hline
ICMP     & $<$256 & o & o & - & + & + & +  & + & + & - & + & + \\ \hline
ISP Traceback & ISP & - & o & - & + & + & +  & + & + & + & + & o \\ \hline\hline
\textbf{Tracemax} & \textbf{$>$50} & \textbf{+} & \textbf{+} & \textbf{+} & \textbf{+} & \textbf{+} & \textbf{+} & \textbf{o} & \textbf{+} & \textbf{+} & \textbf{+} & \textbf{+} \\ \hline
\end{tabular}
\label{tab:comparison}
\vspace{-6mm}
\end{table}

\vspace{1mm}
\section{Conclusion}


We presented the concept of a strong traceback strategy \textit{Tracemax} and an idea to defense against DDoS attacks. It can track the path of just a single packet through the network and address forensic intentions. \textit{Tracemax} can trace significant longer paths than existing methods, detect variable routes during a communication, and distinguish multiple attackers. The technique does not affect the payload data of the packet. 
Therefore, the labeling information does not need to be cleared before delivery. The novel strategy reduces the impact of an attack on the victim and the entire network. The presented solution focus directly on the unwanted traffic with negligible influence on good users. Due to its minimal additional overhead and simple requirements the system results in good scalability and allow a preventive usage. 



\section*{Acknowledgment}
This work was partly funded by ''FLAMINGO'', a Network of Excellence project (ICT-318488) supported by the European Commission under its Seventh Framework program.

\bibliographystyle{IEEEtranS}
\bibliography{IEEEabrv,literature}

\end{document}